\documentclass[aps,prd,preprint,groupedaddress,nofootinbib,showpacs]{revtex4}

\def\lb{\lbrack}
\def\rb{\rbrack}

\def\e{\epsilon}
\def\gg5{\gamma_5}
\def\hg5{\hat{\gamma}_5}
\def\g4{\gamma_4}

\def\Qlatmr1{Q_{lat}^{(m=r=1)}}

\def\be{\begin{eqnarray}}
\def\ee{\end{eqnarray}}

\def\tk{\tilde{k}}

\def\hk{\hat{k}}
\def\hd{\hat{d}}
\def\ha{\hat{a}}

\begin{document}
 
\title{Structure of logarithmically divergent one-loop lattice Feynman 
integrals}

\author{David H. Adams}
\email{dadams@phya.snu.ac.kr}

\author{Weonjong Lee}
\email{wlee@phya.snu.ac.kr}

\affiliation{
Frontier Physics Research Division and Center for Theoretical Physics,
Department of Physics and Astronomy, Seoul National University,
Seoul, 151-747, South Korea}

\date{Sept.~24, 2007}

\begin{abstract}

For logarithmically divergent one-loop lattice Feynman integrals $I(p,a)$, subject to mild 
general conditions, we prove the following expected and crucial structural result:
$I(p,a)=f(p)\log(aM)+g(p)+h(p,M)$
up to terms which vanish for lattice spacing $a\to0$. Here $p$ denotes 
collectively the external momenta and $M$ is a mass scale which may be chosen arbitrarily.
The $f(p)$ and $h(p,M)$ are shown to be universal and coincide with analogous 
quantities in the corresponding continuum integral when the latter is regularized 
either by momentum cut-off or dimensional regularization. The non-universal term
$g(p)$ is shown to be a homogeneous polynomial in $p$ of the same degree as 
$f(p)$. This structure is essential for consistency between renormalized lattice 
and continuum formulations of QCD at one loop.

\end{abstract}

\pacs{11.15.Ha, 11.25.Db}

\maketitle

\section{Introduction}

Logarithmically divergent lattice Feynman integrals are of central importance in 
lattice QCD. The perturbative renormalization factors for the fields, 
bare parameters, and, in many cases, operators of interest, are determined by
such integrals,\footnote{An exception is the additive mass renormalization when chiral symmetry 
is broken, which is given by a linearly divergent lattice integral.} 
and these determine in turn important quantities such as the perturbative 
quantum effective action \cite{R(NPB)}, the beta-function,
the ratio of the lattice and continuum $\Lambda$ parameters, 
and lattice-continuum matching factors for renormalized operators --
see, e.g., \cite{Sachrajda,Capitani} for reviews of lattice perturbation theory and its 
role in extracting physical predictions from lattice QCD.

In this paper we consider logarithmically divergent one-loop lattice 
integrals. Explicit evaluations of such integrals in the past have always resulted in expressions 
which can be written in the form
\be
I(p,a)=f(p)\log(aM)+g(p)+h(p,M)
\label{1.1}
\ee
up to terms which vanish for $a\to0$.
Here $a$ is the lattice spacing, $p$ denotes collectively the external momenta, 
and $M$ is some mass scale (e.g., it can be a fermion mass or the
mass scale of the momentum subtraction renormalization scheme). In particular,
the lattice spacing dependence is given exclusively by the $\log(aM)$ term (there are no
terms $\sim(\log(aM))^{1/3}$ or the like). 
This form is expected \cite{Symanzik,Luscher-Weisz(pert)}, but before now there has been no 
rigorous general proof that the integral must always have this form. Furthermore, for reasons 
discussed below, it is expected that the factor $f(p)$ should be universal, i.e independent of the 
details of the lattice formulation, and that the $a$-independent part of $I(p,a)$ should be given
by the sum of a universal term $h(p,M)$ and a non-universal term $g(p)$ where the latter is a 
homogeneous polynomial in the components of $p$ of the same degree as $f(p)$. We are going to 
prove all of these things in this paper under mild general conditions on $I(p,a)$.

The structure (\ref{1.1}) mirrors the structure of the corresponding continuum integral with some 
choice of regularization:
\be
I^{(c)}(p,\e)=f^{(c)}(p)\,\mbox{div}(\e,M)+g^{(c)}(p)+h^{(c)}(p,M)
\label{1.2}
\ee
up to terms which vanish for $\e\to0$,
where $\e$ denotes the regularization parameter and $\mbox{div}(\e,M)$ is a function that diverges
``logarithmically'' for $\e\to0$. For example, if the regularization is by momentum cut-off 
$\Lambda$ then $\e=1/\Lambda$ and $\mbox{div}(\e,M))=\log(M/\Lambda)$,   
while for dimensional regularization $\mbox{div}(\e,M)=-(lM)^{-\e}/\e$ with $d=4-\e$ 
being the analytic continuation of the spacetime dimension.\footnote{Here $l$ an 
inverse-mass parameter introduced into the integration measure of the Feynman integral by 
$d^4k\to l^{-\e}d^{4-\e}k$ for dimensional reasons.}
Moreover, in explicit evaluations it has always turned out that
\be
f(p)&=&f^{(c)}(p) \label{1.3} \\
h(p,M)&=&h^{(c)}(p,M) \label{1.4} 
\ee 
See, e.g., \cite{Kawai}. This is also expected: 
the relation (\ref{1.3}) must hold in order for the one-loop lattice QCD beta-function to 
coincide with the continuum one, and for the anomalous dimensions of renormalized lattice and 
continuum operators to coincide at one loop as they should. We are going to give a general 
derivation of (\ref{1.3})--(\ref{1.4}) in this paper in the case where the continuum integral
is regularized either by momentum cut-off or dimensional regularization, thereby confirming that 
$f(p)$ and $h(p,M)$ are universal as claimed.

As mentioned, $g(p)$ represents a non-universal term in (\ref{1.1}). Likewise, $g^{(c)}(p)$ 
represents a term in (\ref{1.2}) which is non-universal in the sense that it depends on the
choice of continuum regularization. E.g., the $g^{(c)}(p)$ is generally different for momentum
cut-off and dimensional regularization, while $h^{(c)}(p,M)$ is universal. However, as we will
show, $g(p)$ and $g^{(c)}(p)$ have a relatively simple structure: they are both
homogeneous polynomials in $p$ of the same degree as the homogeneous polynomial $f(p)$.

In practice it often happens that $g(p)$ is proportional to $f(p)$, i.e. $g(p)=cf(p)$, and 
that a factor $f(p)$ can be extracted from $h(p,M)$ leaving a function $\hat{h}(p/M)$.
This is the expected situation for the logarithmically divergent lattice integrals
for the one loop 1PI Greens functions of interest in lattice QCD due to lattice BRST symmetry 
and lattice hypercubic symmetries. Then the integrals can be written in the form
\be
I(p,a)=f(p)\Big(\log(aM)+c+\hat{h}(p/M)\Big),
\label{1.5}
\ee
where only the constant $c$ is non-universal.
When a gauge-invariant regularization (e.g. dimensional regularization) is employed, the 
corresponding continuum Green's function is expressible in a similar way as
\be
I^{(c)}(p,\e)=f(p)\Big(\mbox{div}(\e,M)+c^{(c)}+\hat{h}(p/M)\Big)
\label{1.6}
\ee
Given the continuum structure (\ref{1.6}), the lattice structure (\ref{1.5}) is 
in fact crucial for consistency between lattice and continuum formulations of QCD at 
one loop. We intend to prove (\ref{1.5}) for the one loop 1PI Green's functions for 
general lattice formulations of QCD in future work.
The results of the present paper are clearly an essential step in this direction. 

Our results in this paper are specific to the one loop case. The higher loop case involves 
additional technical complications and is left for future work. 

In the statement of results
above we have not indicated the dependence on mass parameters, e.g. fermion masses 
(if there are any present). However, this dependence is easily described: 
In the derivations of the results that we give, mass parameters 
enter in an analogous way to the external momenta, so the dependence on them is given by simply 
replacing $p\to(p,m)$ in the statements of the results above, where $m$ denotes collectively
all the mass parameters. Or we can simply take $p$ to denote collectively all the external 
momenta {\em and} masses, which is what we will do in the subsequent sections. 


The main steps in our derivation of the results are as follows.
We begin by separating out a divergent part $I_0(p,M,a)$ of $I(p,a)$; it is essentially just
the leading term in the Taylor-expansion of $I(p,a)$ in $p$ with a mass parameter $M$ introduced 
to regulate the infrared divergence. The difference $I(p,a)-I_0(p,M,a)$ is shown, as expected,
to be expressible as a convergent lattice integral whose $a\to0$ limit we denote by $h(p,M)$ 
(it is seen to coincide with the analogous continuum quantity $h^{(c)}(p,M)$). 
Then we show that $f(p):=\lim_{a\to0}\;a\frac{d}{da}I_0(p,M,a)$ is finite, independent of $M$, 
and given by a convergent continuum integral which is shown to coincide with $f^{(c)}(p)$. 
The final, and most technically challenging, step is to show that 
$g(p):=\lim_{a\to0}\;(\,I_0(p,M,a)-f(p)\log(aM))$ is finite 
and independent of $M$. Altogether this implies (\ref{1.1}) with (\ref{1.3})--(\ref{1.4}).
From its construction it will be clear that $I_0(p,M,a)$ is a homogeneous polynomial in $p$,
which implies that $f(p)$ and $g(p)$ are also homogeneous polynomials of the same degree.

The derivations of the results regarding $f(p)$, $g(p)$ and 
$h(p,M)$ involve applications of Reisz's lattice power-counting theorem 
\cite{R,R(massless)}. (The power-counting theorem does not apply to the logarithmically
divergent lattice integral $I(p,a)$ but we apply it to certain convergent integrals
associated with $I(p,a)$ to establish the desired results.) Furthermore, to establish the 
finiteness of $g(p)$ we require an {\em extension} of the power-counting theorem, namely a
bound on how quickly a convergent lattice integral approaches its continuum limit
for $a\to0$. We derive this extension in the present paper, and in the process give a proof of
the power-counting theorem for one-loop lattice integrals which does not require a certain
technical condition on the propagators that was needed in Reisz's general proof. This allows
us to establish the present structural results under milder conditions on the lattice
integral than in Reisz's work.

The organization of this paper is as follows. In \S2 we describe the general lattice integrals 
that we consider and state the mild, general conditions on them that we require to derive the 
advertised results. 
In \S3 we introduce $I_0(p,M,a)$ and derived the mentioned result regarding $h(p,M)$.
Then in \S4 we derive the results regarding $f(p)$ and $g(p)$, thereby establishing the 
general structural results for $I(p,a)$ discussed above. We check the results in an 
illustrative example in \S5, and conclude in \S6 with a summary and
discussion of possibilities for extending the results of the paper beyond one loop.
The paper contains several appendices. 
In Appendix A we prove the extension of the one-loop lattice power-counting theorem used in the 
derivation of the structural results of this paper. In Appendix B 
we show $f(p)=f^{(c)}(p)$ in the case of dimensional regularization of the continuum integral.
(The proof of this in the case of momentum cut-off is given in \S4.)

Some of the techniques and results of this paper were developed previously in a special case
in Ref.\cite{prev}

\section{The setup}

The one-loop lattice Feynman integrals that we consider have the general form 
\be
I(p,a)=\int_{-\pi/a}^{\pi/a}d^4k\,\frac{V(k,p,a)}{C(k,p,a)}
\label{2.1}
\ee
where $p$ denotes collectively all the external momenta and masses. 
The functions in the integrand in (\ref{2.1}) have the general form
\be
V(k,p,a)=\frac{1}{a^m}F(ak,ap)\quad,\qquad C(k,p,a)=\frac{1}{a^n}G(ak,ap)
\label{2.2}
\ee
where $F$ and $G$ are smooth functions. The lattice degree of $V(k,p,a)$ as a function of $k$
\cite{R} can be characterized as follows. Let $r$ be the order of the first non-vanishing term 
in the Taylor expansion of $F(ak,tap)$ around $t=0$, then
\be
F(ak,tap)=t^rF_0(ak,ap)+t^{r+1}F_1(ak,ap,t)
\label{2.3}
\ee
where $t^rF_0(ak,ap)$ is the first non-vanishing term in the Taylor expansion and $F_1(k,p,t)$ is 
a smooth function; in particular it and its derivatives are finite at $t=0$. It follows from
(\ref{2.2})--(\ref{2.3}) that
\be
V(\lambda k,p,\frac{a}{\lambda})\ \sim\ \lambda^{m-r}\qquad\mbox{for $\lambda\to\infty$},
\label{2.4}
\ee
hence the lattice degree of $V(k,p,a)$ in $k$, which we denote by $\hd_V$, is $m-r$ \cite{R}.

We require that the first non-vanishing term in the expansion of $G(ak,tap)$ around $t=0$ is the
zero-order term; then the expansion has the form
\be
G(ak,tap)=G(ak,0)+tG_1(ak,ap,t)
\label{2.4a}
\ee
where $G(ak,ap,t)$ is smooth including at $t=0$. It follows from this and (\ref{2.2}) that
\be
C(\lambda k,p,\frac{a}{\lambda})\ \sim\ \lambda^n\qquad\mbox{for $\lambda\to\infty$},
\label{2.4b}
\ee
hence the lattice degree of $C(k,p,a)$ in $k$ is $\hd_C=n$.
The divergence degree of the lattice integral is then defined as \cite{R}
\be
\hd_I:=4+\hd_V-\hd_C=4+m-r-n\,.
\label{2.5}
\ee
Henceforth we assume that $V(k,p,a)$ and $C(k,p,a)$ have finite continuum limits,
\be
V(k,p,a)\stackrel{a\to0}{\longrightarrow}P(k,p)\quad,\qquad 
C(k,p,a)\stackrel{a\to0}{\longrightarrow}E(k,p)\,.
\label{2.6}
\ee
Then, in light of (\ref{2.2}), $P(k,p)$ and $E(k,p)$ are homogeneous polynomials 
in $(k,p)$ of degree $m$ and $n$, respectively. We denote the usual polynomial degrees of
$P(k,p)$ and $E(k,p)$ in $k$ by $d_P$ and $d_E$, respectively. 
As pointed out in Eq.~(2-11) of \cite{R} the lattice degree of $C$ and continuum degree of $P$ 
need not be equal, but the inequality $d_P\le\hd_V$ holds. However, in the case of $P$ and $E$,
the condition (ii) imposed on $E$ below ensures that $d_E=n=\hd_P$.

Further conditions to be imposed on the lattice integral are the following:

\noindent (i) $G(ak,ap)\ge0$ $\,\forall k,p,a$. (Then $C(k,p,a)$ and $E(k,p)$ are also positive 
functions).

\noindent (ii) $E(k,p)\;\sim\;|k|^n$ for $|k|\to\infty$.\footnote{Since $E(k,p)$ is a polynomial,
this implies that $n$ must be even so that $|k|^n=(k^2)^{n'}$ for integer $n'=n/2$.}

\noindent (iii) $G(ak,0)\ne0$ for all non-zero $k\in[-\pi/a,\pi/a]^4$ (the ``doubler-free''
condition).\footnote{Note that $G(ak,0)$ 
necessarily vanishes at $k\!=\!0$ since otherwise $C(k,p,a)=\frac{1}{a^m}G(ak,ap)$ 
could not have a finite $a\to0$ limit.}

\noindent In practice the denominator function $C(k,p,a)$ 
arises as a product of lattice propagators
in such a way that conditions (i)--(ii) are automatically satisfied. This was the situation 
considered by Reisz in his proof of the lattice power-counting theorem \cite{R,R(massless)}.
However, to prove the theorem Reisz required an 
additional technical condition on the lattice propagators -- see \S10.1 of Ref.\cite{Capitani} 
for a discussion (the condition is denoted there by ``(C3)''). 
This needs to be established for each lattice formulation that one considers, and it can 
sometimes be non-trivial. We are able to make do without this additional condition in this 
paper, since we are able to prove the power-counting theorem for one loop lattice integrals
in Appendix A using only the conditions stated above (and the infrared finiteness condition
mentioned below).

In Reisz's initial derivation of the lattice power-counting theorem the propagators were required
to be massive \cite{R}, corresponding to strictly positive functions in condition (i) above.
He subsequently extended the theorem to allow for massless propagators in Ref.\cite{R(massless)} 
where the notion of infrared divergence degree was introduced to handle them.
We also require here that the infrared divergence degree of the lattice integral be strictly 
negative so that the integral is infrared finite. In practice, for one loop lattice integrals
with massless propagators, this usually means that the external momenta must be non-vanishing. 

The continuum version of the lattice integral (\ref{2.1}) is
\be
I^{(c)}(p)=\int_{-\infty}^{\infty}d^4k\,\frac{P(k,p)}{E(k,p)}
\label{2.10x}
\ee
The one-loop lattice power-counting theorem (Appendix A) states that when $\hd_I<0$ and the 
lattice integrand satisfies
the aforementioned conditions then the $a\to0$ limit of $I(p,a)$ is finite and coincides with
$I^{(c)}(p)$. The theorem cannot be applied directly to logarithmically divergent ($\hd_I=0$) 
lattice integrals, but we will apply it to certain convergent integral associated with them
to establish the structural results of this paper. We will also need an extension
of the power counting theorem, namely a bound on how quickly $I(p,a)$ converges to $I^{(c)}(p)$
when $\hd_I<0$, which we derive in Appendix A. It tells that $I(p,a)-I^{(c)}(p)$ vanishes
at least as quickly as $\sim a\log(1/a)$ for $a\to0$ -- see Appendix A for the precise statement.

\section{Splitting off the divergent part}

We now specialize to logarithmically divergent ($\hd_I=0$) one loop 
lattice integrals $I(p,a)$ satisfying the conditions of \S2. To study such integrals 
it is useful to split them into the sum of a simpler logarithmically divergent integral 
$I_0(p,M,a)$ and convergent integral $I_1(p,M,a)=I(p,a)-I_0(p,M,a)$ as described in this section.

In light of (\ref{2.3}) we can decompose
\be 
V(k,p,a)=V_0(k,p,a)+V_1(k,p,a)
\label{3z.1}
\ee
where
\be
V_0(k,p,a)&=&\frac{1}{a^m}F_0(ak,ap)\ \ ,\quad\ \hd_{V_0}=\hd_V \label{3z.2} \\ 
V_1(k,p,a)&=&\frac{1}{a^m}F_1(ak,ap,1)\ \ ,\quad \hd_{V_1}\,\le\,\hd_V-1 \label{3z.3}
\ee
The statements regarding $\hd_{V_0}$ and $\hd_{V_1}$ in (\ref{3z.2})--(\ref{3z.3}) are
seen as follows. Since $t^rF_0(ak,ap)$ is the order $r$ term in the expansion of $F(ak,tap)$ 
in $t$ it is a homogeneous polynomial in $p$ of degree $r$. Hence,
\be
F_0(ak,tap)=t^rF_0(ak,ap)
\label{3z.4}
\ee
and it follows that $V_0(\lambda k,p,\frac{a}{\lambda})\sim\lambda^{m-r}$ for $\lambda\to\infty$.
Next, using (\ref{2.3}) we find $F_1(ak,tap,1)=F(ak,tap)-F_0(ak,tap)$ and then, 
again using (\ref{2.3}),
\be
F_1(ak,tap,1)=t^{r+1}F_1(ak,ap,t)\,.
\label{3z.5}
\ee
It follows that
$V_1(\lambda k,p,\frac{a}{\lambda})=\frac{\lambda^{m-r-1}}{a^m}F_1(ak,ap,1/\lambda)$ which 
diverges no quicker than $\sim\lambda^{m-r-1}$ for $\lambda\to\infty$.

Now let $M$ be an arbitrary mass parameter and define\footnote{Note that 
$C(k,0,a)+M^n=\frac{1}{a^n}\widetilde{G}(ak,aM)$ with $\widetilde{G}(ak,aM)=G(ak,0)+(aM)^n$, 
so this integral is of the form considered in \S2.}
\be
I_0(p,M,a)=\int_{-\pi/a}^{\pi/a}d^4k\,\frac{V_0(k,p,a)}{C(k,0,a)+M^n}
\label{3z.6}
\ee
Since $G(ak,0)$ is non-vanishing, $C(\lambda k,0,\frac{a}{\lambda})\sim\lambda^n$ for 
$\lambda\to\infty$ just as in (\ref{2.4b}), hence the divergence degree of the denominator function
here is $n$ just as before. This together with (\ref{3z.2}) implies
$\hd_{I_0}=\hd_I=0$. Clearly $I_0(p,0,a)$ is precisely the leading order
term in the Taylor expansion of $I(p,a)$ in the external momenta $p$. It is generally infrared
divergent though, and $I_0(p,M,a)$ is simply an infrared-regularized version of this term with
$M$ being the regulator. ($M$ is raised to the power of $n$ in (\ref{3z.6}) to ensure that
the regulator term $M^n$ has the appropriate mass-dimension.)
In light of this we expect the difference between $I(p,a)$ and $I_0(p,M,a)$ to be a convergent 
lattice integral, and this will be explicitly verified below. 
Therefore the structure of $I(p,a)$ can be inferred from that of $I_0(p,M,a)$. 
The latter is easier to study since its dependence on $p$ is simpler -- it enters
only through the numerator $V_0(k,p,a)$ in the integrand. 
By taking the denominator in the integrand to be $C(k,0,a)+M^n$ rather than $C(k,p,a)$ we have
traded the potentially complicated $p$-dependence for a more straightforward dependence
on the mass parameter $M$. 

Note also that since $F_0(ak,ap)$ is a 
homogeneous polynomial of degree $r$ in $p$ the same is true for $V_0(k,p,a)$ and therefore 
also for $I_0(p,M,a)$.

To explicitly verify that $I(p,a)$ and $I_0(p,M,a)$ differ by a convergent integral, we start from
\be
I(p,a)-I_0(p,M,a)=\int_{-\pi/a}^{\pi/a}d^4k\,\bigg\lb\frac{V(k,p,a)}{C(k,p,a)}
-\frac{V_0(k,p,a)}{C(k,0,a)+M^n}\bigg\rb
\label{3z.7}
\ee
and rewrite the integrand as
\be
\frac{V(k,p,a)-V_0(k,p,a)}{C(k,p,a)}-V_0(k,p,a)\bigg\lb\frac{C(k,p,a)-C(k,0,a)-M^n}
{C(k,p,a)(C(k,0,a)+M^n)}\bigg\rb\,.
\label{3z.8}
\ee
The integral of the first term here has divergence degree $<0$ since $V-V_0=V_1$ has divergence
degree $\le \hd_V\!-\!1$ (recall (\ref{3z.3})). The integral of the second term also 
has divergence degree $<0$ since $C(k,p,a)-C(k,0,a)-M^n$ has divergence degree $\le n\!-\!1$;
this follows from the fact that $C(\lambda k,p,a/\lambda)-C(\lambda k,0,a/\lambda)
=\frac{\lambda^{n-1}}{a^n}G_1(ak,ap,1/\lambda)$ where we have used (\ref{2.2}) and (\ref{2.4a}).
The lattice power counting theorem then implies that
(\ref{3z.7}) has a finite $a\to0$ limit as claimed, and that the limit is given by
the corresponding continuum integral:
\be
h(p,M)&:=&\lim_{a\to0}\ \Big(I(p,a)-I_0(p,M,a)\Big) \nonumber \\
&=&\int_{-\infty}^{\infty}d^4k\,\bigg\lb\frac{P(k,p)}{E(k,p)}
-\frac{P_0(k,p)}{E(k,0)+M^n}\bigg\rb
\label{3z.10}
\ee
with 
\be
P_0(k,p):=\lim_{a\to0}\ V_0(k,p,a)
\label{3z.11}
\ee
$P_0(k,p)$ is the term of order $r$ in the (finite) expansion of $P(k,p)$ in powers of $p$. 
It can happen that $P_0$ vanishes; this is the case when $d_P<\hd_V$. Non-vanishing
$P_0$ corresponds to $d_P=\hd_V$. In the latter case both $I^{(c)}(p)$ and $I_0^{(c)}(p,M)$ 
are logarithmically divergent and need to be regularized; here $I_0^{(c)}(p,M)$ denotes
the continuum version of $I_0(p,M,a)$ given by
\be
I_0^{(c)}(p,M)=\int_{-\infty}^{\infty}d^4k\,\frac{P_0(k,p)}{E(k,0)+M^n}\,.
\label{3z.12}
\ee
However, the difference
$I^{(c)}(p)-I_0^{(c)}(p,M)$ can be written as a convergent integral just as in the
lattice case above after invoking the continuum power counting theorem, and this integral is
precisely the one in (\ref{3z.10}). By definition $h^{(c)}(p,M)$ is the difference
$I^{(c)}(p)-I_0^{(c)}(p,M)$ in the limit where the continuum regularization is lifted, so we 
have verified that $h(p,M)=h^{(c)}(p,M)$ as claimed in (\ref{1.4}) in the Introduction.

Regarding the $M$-dependence of $h(p,M)$ we note the following. After changing variables to
$\hk=k/M$ in (\ref{3z.10}) and using the aforementioned facts that $P(k,p)$ and $E(k,p)$ are 
homogeneous polynomials in $(k,p)$ of orders $m$ and $n$, respectively, we find
\be
h(p,M)=\int_{-\infty}^{\infty}d^4\hk\,M^r\bigg\lb\frac{P(\hk,p/M)}{E(\hk,p/M)}
-\frac{P_0(\hk,p/M)}{E(\hk,0)+1}\bigg\rb
\label{3z.13}
\ee
Since the lowest order term in the expansion of $P(k,p)$ in $p$ has order $\ge r$ it follows that
$h(p,M)$ has the general form
\be
h(p,M)=\sum_{r_1+\dots+r_N=r}p_1^{r_1}\cdots p_N^{r_N}\hat{h}_{r_1\cdots r_N}(p/M)
\label{3z.14}
\ee
where $\{p_1,\dots,p_N\}$ denotes all the components of $p$.
As mentioned in the introduction, it often happens in practice that $h(p,M)$ has the form 
$f(p)\hat{h}(p/M)$, which is a special case of (\ref{3z.14}). 
However, it does not seem possible to 
derive this form on general grounds without requiring further properties of the lattice intergal
(e.g., resulting from lattice BRST symmetry and lattice hypercube symmetries).

So far we have essentially been following the usual procedure for studying a logarithmically
divergent Feynman integral by subtracting off (a suitably regularized version of) the leading
term in the momentum expansion. The main new content of this paper comes in the next section
where we derive the general structural results for this term.

\section{Structure of $I_0(p,M,a)$}

Our goal in this section is to show that $I_0(p,M,a)$ has the general form
\be
I_0(p,M,a)=f(p)\log(aM)+g(p)
\label{3.1}
\ee
up to terms which vanish for $a\to0$, where $f(p)$ is given by a convergent continuum integral
(and hence is universal) and coincides with the analogous factor $f^{(c)}(p)$ in the expression
for the corresponding continuum integral $I_0^{(c)}(p,M)$ when the latter is regularized either
by momentum cut-off or dimensional regularization. Note that since $I_0(p,M,a)$ is a homogeneous
polynomial in $p$ of degree $r$ the same must then be true for $f(p)$ and $g(p)$. When combined 
with the results in the preceding section this then implies the general structural results
(\ref{1.1}) and (\ref{1.3})--(\ref{1.4}) for $I(p,a)$ stated in the Introduction.

The demonstration proceeds in two steps. First we define
\be
f(p):=\lim_{a\to0}\ a\frac{d}{da}I_0(p,M,a)
\label{3.2}
\ee
and show that this limit is finite, independent of $M$,  
and given by a convergent continuum integral which coincides with $f^{(c)}(p)$. Then we show
that the limit 
\be
g(p):=\lim_{a\to0}\ \big(I_0(p,M,a)-f(p)\log(aM)\big)
\label{3.3}
\ee
is finite and independent of $M$. 
This implies that (\ref{3.1}) holds up to terms which vanish for $a\to0$ as claimed. 
Both of these steps require the lattice power-counting theorem; the second step requires the
extension of the power counting theorem which we prove in Appendix A.

A change of variables in (\ref{3z.6}) leads to 
\be
I_0(p,M,a)=\int_{-\pi}^{\pi}d^4\hk\,\frac{F_0(\hk,p)}{G(\hk,0)+(aM)^n}
\label{3.4}
\ee
where we have used the facts that $\hd_I=4+m-r-n=0$ and $F_0(ak,ap)$ is a homogeneous polynomial
in $p$ of degree $r$. Note that $F_0(\hk,p)$ is dimensionful in (\ref{3.4}) with mass-dimension 
$r$ since $p$ is dimensionful while $\hk$ is dimensionless. Setting
\be
\ha=aM
\label{3.5}
\ee
we find from (\ref{3.4}) that
\be
a\frac{a}{da}\,I_0(p,M,a)
&=&-n\ha^n\int_{-\pi}^{\pi}d^4\hk\,\frac{F_0(\hk,p)}{(G(\hk,0)+\ha^n)^2} \nonumber \\
&=&-n\int_{-\pi/\ha}^{\pi/\ha}d^4\hk\,\frac{\frac{1}{\ha^m}F_0(\ha\hk,\ha p)}
{(\frac{1}{\ha^n}G(\ha\hk,0)+1)^2}
\label{3.6}
\ee
The lattice divergence degrees of the numerator and denominator functions here are $m-r$ and
$2n$, respectively, so the integral has divergence degree $4+m-r-2n=-n<0$. We conclude from the 
power counting theorem that its $a\to0$ limit is finite and given by
\be
f(p)=-n\int_{-\infty}^{\infty}d^4\hk\,\frac{P_0(\hk,p)}{(E(\hk,0)+1)^2}
\label{3.7}
\ee
independent of $M$.

We now show that $f(p)=f^{(c)}(p)$ when the continuum integral is regularized by momentum
cut-off. With this regularization the continuum $I_0$-integral,
\be
I_0^{(c)}(p,M,\Lambda)=\int_{-\Lambda}^{\Lambda}d^4k\,\frac{P_0(k,p)}{E(k,0)+M^n}\,,
\label{3n.1}
\ee
has the structure
\be
I_0^{(c)}(p,M,\Lambda)=f^{(c)}(p)\log(M/\Lambda)+g^{(c)}(p)
\label{3n.2}
\ee
up to terms which vanish for $\Lambda\to\infty$. (This can be seen by an analogue of the argument
we give in the present lattice case; the continuum argument is simpler and we omit it.)
This implies the structure (\ref{1.2}) for the continuum integral with momentum cut-off 
regularization discussed in the Introduction, since $I^{(c)}(p,\Lambda)-I_0^{(c)}(p,M,\Lambda)$ 
reduces to the convergent integral $h^{(c)}(p,M)$ for $\Lambda\to\infty$ (cf. \S3). 
Moreover, we see that   
\be
f^{(c)}(p)=\lim_{\Lambda\to\infty}\ -\Lambda\frac{d}{d\Lambda}I_0^{(c)}(p,M,\Lambda)\,.
\label{3.8}
\ee
Changing variables to $\hk=k/\Lambda$ in (\ref{3n.1}) leads to
\be
I_0^{(c)}(p,M,\Lambda)=\int_{-1}^{1}d^4\hk\,\frac{P_0(\hk,p)}{E(\hk,0)+(\frac{M}{\Lambda})^n}
\label{3.9}
\ee
Starting from this, (\ref{3.8}) is easily evaluated similarly to the lattice case, 
and is found to reproduce the integral (\ref{3.7}) 
for $f(p)$.\footnote{The calculation involves a change of variables and exploits the facts 
that $P_0(k,p)$ is homogeneous of degree $m-r$ in $k$ (since, as noted previously, it must be 
homogeneous of degree $m$ in $(k,p)$ while also homogeneous of degree $r$ in $p$) and $E(k,0)$
is homogeneous of degree $n$ in $k$ (since as already noted $E(k,p)$ must be homogeneous
of degree $n$ in $(k,p)$).} The proof of $f(p)=f^{(c)}(p)$ in the case of dimensional
regularization is given in Appendix B.

To accomplish the remaining step -- to prove that the limit in (\ref{3.3}) is finite --
we start by noting from (\ref{3.4}) that $I_0(p,M,a)$ depends on $M$ and $a$ through the product
$\ha=aM$. We will therefore also denote the integral by $I_0(p,\ha)$ in the following.
Define
\be
f(p,\ha):=a\frac{d}{da}I_0(p,M,a)=\ha\frac{d}{d\ha}I_0(p,\ha)\,,
\label{3.9a}
\ee
then $f(p,\ha)\to f(p)$ for $\ha\to0$ and the quantity in (\ref{3.3}) can be expressed as
\be
I_0(p,\ha)-f(p)\log(\ha)=-\int_{\ha}^1db\,\frac{1}{b}\Big(f(p,b)-f(p)\Big)+I_0(p,1)
\label{3.10}
\ee
To show that this has finite $a\to0$ limit, or equivalently, finite $\ha\to0$ limit,
we need to show that the integral on the right-hand side remains finite for $\ha\to0$.
For this we need information on how quickly $f(p,b)$ approaches its continuum limit $f(p)$ 
for $b\to0$. The $f(p,b)$ has the lattice integral expression given by (\ref{3.6}) with 
$\ha$ replaced by $b$, and we noted there that that integral has strictly negative divergence 
degree. The extension of the lattice power counting theorem proved in Appendix A then tells 
that $f(p,b)-f(p)$ vanishes at least as fast as $\sim b\log(1/b)$ for $b\to0$. 
It follows that the integral
\be
\int_{\ha}^1db\,\frac{1}{b}|f(p,b)-f(p)|
\label{3.11}
\ee
remains finite in the $\ha\to0$ limit. By Lebesgue's ``theorem of dominated convergence''
the integral continues to have a well-defined finite limit when the integrand is replaced by 
$\frac{1}{b}(f(p,b)-f(p))$. Hence the $\ha\to0$ limit of (\ref{3.10}) is finite. Since $M$ 
only enters there through $\ha$ there is no $M$-dependence remaining in the $\ha\to0$ limit;
i.e., $g(p)$ is both finite and independent of $M$ as claimed.
This completes the demonstration of the general structural result (\ref{3.1}), thereby 
establishing the main results (\ref{1.1}) and (\ref{1.3})--(\ref{1.4}) of this paper.

Although $g(p)$ arises as the $a\to0$ limit of a lattice expression it is non-universal in general.
For example, in the logarithmically divergent lattice integral expression for the gluonic
2-point function at one loop in lattice QCD with Wilson fermions, $g(p)$ depends on the Wilson
parameter, cf. Eq.(3.24)--(3.25) of Ref.\cite{Kawai}.

We remark that since $f(p)$ and $g(p)$ are homogeneous polynomials of order $r$, and since
$h(p,M)$ has the form (\ref{3z.14}), the general expression (\ref{1.1}) for the lattice integral
can be written as
\be
I(p,a)=\sum_{r_1+\dots+r_N=r}p_1^{r_1}\cdots p_N^{r_N}\Big(f_{r_1\cdots r_N}\log(aM)
+g_{r_1\cdots r_N}+\hat{h}_{r_1\cdots r_N}(p/M)
\label{3.12}\Big)
\ee
where the coefficients $f_{r_1\cdots r_N}$ and functions $\hat{h}_{r_1\cdots r_N}(p/M)$ are 
universal (being determined by $f(p)$ and $h(p,M)$, respectively) while the coefficients
$g_{r_1\cdots r_N}$ are non-universal. (The expression (\ref{1.5}) mentioned in the Introduction
is a special case of this.)

\section{An illustrative example}

As a check on the results of this paper, and
illustration of how they can be used in practice, we apply them
to the following lattice integral which appeared in Eq.(3.9) of Ref.\cite{Kawai} in connection
with the ghost self-energy at one loop in lattice QCD:
\be
I(p,a)_{\mu}=\int_{-\pi/a}^{\pi/a}\frac{d^4k}{(2\pi)^4}\,
\frac{\frac{2}{a}\sin\frac{a}{2}(k+p)_{\mu}\cos\frac{a}{2}(k+p)_{\mu}}
{\Big(\sum_{\sigma}\frac{4}{a^2}\sin^2\frac{a}{2}k_{\sigma}\Big)
\Big(\sum_{\rho}\frac{4}{a^2}\sin^2\frac{a}{2}(k+p)_{\rho}\Big)}
\label{5.1}
\ee
Evaluation of this integral in Ref.\cite{Kawai} involved determining the leading order term
in the expansion in $p$ and evaluating the logarithmically divergent lattice integral expression 
for this term, using dimensional regularization to deal with its infrared divergence.
The results of the present paper allow to evaluate (\ref{5.1}) entirely by continuum integral
calculations, except for a non-universal constant (which in practice has to be determined
numerically anyway). 

As it stands, the integral (\ref{5.1}) appears to be linearly divergent. It is actually
logarithmically divergent though, as becomes manifest after symmetrizing the integrand under
$k\to-k$. (Or equivalently, exploiting the symmetry $I_{\mu}(p,a)=-I_{\mu}(-p,a)$ to write
$I(p,a)_{\mu}=\frac{1}{2}(I(p,a)_{\mu}-I(-p,a)_{\mu})$.) The integral is then expressed as
\be
I(p,a)_{\mu}=\frac{1}{2}\int_{-\pi/a}^{\pi/a}\frac{d^4k}{(2\pi)^4}\,
\frac{V(k,p,a)}{C(k,p,a)}
\label{5.2}
\ee
where
\be
V(k,p,a)=\frac{1}{a^3}F(ak,ap)\quad,\qquad C(k,p,a)=\frac{1}{a^6}G(ak,ap)
\label{5.3}
\ee
with 
\be
F(k,p)&=&2\sin{\textstyle \frac{1}{2}}(k+p)_{\mu}\cos{\textstyle \frac{1}{2}}(k+p)_{\mu}
\Big(\sum_{\sigma}4\sin^2{\textstyle \frac{1}{2}}(k-p)_{\rho}\Big) \nonumber \\
&&-\ 2\sin{\textstyle \frac{1}{2}}(k-p)_{\mu}\cos{\textstyle \frac{1}{2}}(k-p)_{\mu}
\Big(\sum_{\sigma}4\sin^2{\textstyle \frac{1}{2}}(k+p)_{\rho}\Big) \label{5.4} \\
G(k,p)&=&
\Big(\sum_{\sigma}4\sin^2{\textstyle \frac{1}{2}}k_{\sigma}\Big)
\Big(\sum_{\rho}4\sin^2{\textstyle \frac{1}{2}}(k+p)_{\rho}\Big)
\Big(\sum_{\rho}4\sin^2{\textstyle \frac{1}{2}}(k-p)_{\rho}\Big) \label{5.5}
\ee
This is a particular case of our general setting with $m\!=\!3$, $n\!=\!6$ and $r\!=\!1$
(note that $F(k,0)=0$; the first non-vanishing term in the expansion of $F(k,p)$ in $p$ is the 
linear one). The ingredients that we need for the continuum integrals are now easily 
found: 
\be
P(k,p)=\lim_{a\to0}\;V(k,p,a)&=&(k+p)_{\mu}(k-p)^2-(k-p)_{\mu}(k+p)^2 \nonumber \\
&=&2p_{\mu}k^2-4k_{\mu}(kp)+2p_{\mu}p^2 \label{5.6} \\ 
E(k,p)=\lim_{a\to0}\;C(k,p,a)&=&k^2(k+p)^2(k-p)^2 \nonumber \\
&=&k^2((k^2+p^2)^2-4(kp)^2)
\label{5.7}
\ee
which imply
\be
P_0(k,p)&=&2p_{\mu}k^2-4k_{\mu}(kp)
 \label{5.8} \\
E(k,0)&=&(k^2)^3 \label{5.9}
\ee
Recall that the $p$-dependence of $I_0(p,M,a)$ is determined by that of $F_0(k,p)$; it can be
found from (\ref{5.4}) using $F_0(k,p)=\frac{d}{dt}F(k,tp)_{t=0}$. We will not need the explicit 
expression here but simply note the following property: $F_0(k,p)$ has the form 
$\sum_{\sigma}p_{\sigma}F_{0\sigma}(k)$ and $F_{0\sigma}(k)$ changes sign under 
$k_{\sigma}\to-k_{\sigma}$ when $\sigma\ne\mu$; the $p_{\sigma}$-terms are thus seen to give 
vanishing contribution to $I_0(p,M,a)$ when $\sigma\ne\mu$. This together with the general
results of this paper implies 
\be
I(p,a)_{\mu}=p_{\mu}(c_0\log(aM)+c_1)+h(p,M)
\label{5.10}
\ee
(up to terms which vanish for $a\to0$). Here $p_{\mu}c_0=f(p)$; this and $h(p,M)$ are given by the 
continuum integrals (\ref{3.7}) and (\ref{3z.10}), respectively, while $c_1$ is a non-universal 
numerical constant which our results do not determine. In light of (\ref{5.2}) we need to first 
replace $\int d^4k\to\frac{1}{2}\int \frac{d^4k}{(2\pi)^4}$ in the continuum integral expressions,
then the explicit evaluation of $f(p)$ gives
\be
f(p)&=&-\frac{6}{2}\int_{-\infty}^{\infty}\frac{d^4k}{(2\pi)^4}\,
\frac{2p_{\mu}k^2-4k_{\mu}(kp)}{((k^2)^3+1)^2} \nonumber \\
&=&-3\int_{-\infty}^{\infty}\frac{d^4k}{(2\pi)^4}\,\frac{p_{\mu}k^2}{(k^6+1)^2} \nonumber \\
&=&-3p_{\mu}\int_0^{\infty}\frac{dr}{8\pi^2}\,\frac{r^5}{(r^6+1)^2} \nonumber \\
&=&-\frac{1}{16\pi^2}\,p_{\mu} 
\label{5.11}
\ee
hence $c_0=-1/16\pi^2$ in (\ref{5.10}). Turning now to $h(p,M)$, rather than evaluating the
continuum integral (\ref{3z.10}) explicitly we can proceed as follows. In the present case 
$p_{\mu}$ can be factored out in (\ref{3z.10}) leaving an expression that
depends on $p$ only through $|p|$.\footnote{To see this, decompose the integration variable 
into $k=t\frac{p}{|p|}+q$ where $q\perp p$. Then, after noting $k^2=t^2+q^2$, 
$\,kp=t|p|$ and $k_{\mu}(kp)=t^2p_{\mu}+q_{\mu}t|p|$ we see from (\ref{5.6}) that the terms in
$P(k,p)$ contain an overall factor of $p_{\mu}$ except for the term $\sim q_{\mu}$, while from
(\ref{5.7}) $E(k,p)$ is seen to depend on $(t,q)$ only through $t^2$ and $q^2$. It follows that
the term $\sim q_{\mu}$ in $P(k,p)$ gives vanishing contribution to (\ref{3z.10}) since it
changes sign under $q\to-q$. And after $p_{\mu}$ is factored out in the other terms, the remaining
expression manifestly depends on $p$ only through $|p|$ as claimed.}
In light of (\ref{3z.14}) we conclude that $h(p,M)$ has the general form 
$p_{\mu}\hat{h}(\frac{|p|}{M})$.
Setting $M=|p|$ it follows that
\be
I(p,a)_{\mu}=p_{\mu}\Big(-\frac{1}{32\pi^2}\log(a^2p^2)+c\Big)
\label{5.12}
\ee
up to terms which vanish for $a\to0$,
where $c=c_1+\hat{h}(1)$. Thus we have reproduced the result Eq.(3.19b) of Ref.\cite{Kawai}
except for the undetermined numerical constant $c$. The determination of this constant in
Ref.\cite{Kawai} required knowledge of the numerically determined constant appearing in a certain 
logarithmically divergent lattice integral. In the present case, $c$ could be directly determined
numerically by taking $\mu\!=\!1$, setting $p=(1,0,0,0)$ and fitting $I(p,a)$ to the right-hand
side of (\ref{5.12}) in the limit where $a^2p^2$ becomes small and the lattice volume 
becomes large.

\section{Summary and concluding remarks}

The main new technical results of this paper, on which the derivation of the structure result
was based, can be summarized as follows:

\noindent (1) The limit
$f(p):=\lim_{a\to0}\ a\frac{d}{da}I(p,a)$ is finite and universal, given by a convergent
continuum integral (\ref{3.7}) which coincides with the analogous continuum factor
$f^{(c)}(p)$.

\noindent (2) The limit  
$g(p,M):=\lim_{a\to0}\;(\,I(p,a)-f(p)\log(aM))$ is finite.

\noindent More precisely, these results were established for the 
simpler integral $I_0(p,M,a)$ and then hold for $I(p,a)$ as well since it differs from 
$I_0(p,M,a)$ by a convergent lattice integral; this also leads to the decomposition 
$g(p,M)=g(p)+h(p,M)$ with $h(p,M)$
being universal and $g(p)$ a non-universal homogeneous polynomial of the same degree as $f(p)$.
The structural results for $I(p,a)$ stated in the Introduction are essentially consequences of 
(1)--(2) above.

The proof of (1) was relatively straightforward and basically amounted to showing that the 
lattice integral expression for $a\frac{d}{da}I(p,a)$ has strictly negative divergence degree,
so that the lattice power-counting theorem can be applied. Proving (2) is the ``hard part''
of this work; it requires an extension of the lattice power-counting theorem in the one-loop
case, namely a bound on how quickly a convergent lattice integral converges to
its continuum limit, which we have given in Appendix A.

The conditions on the lattice integral required in this work can be summarized by saying that
they are the same as required by Reisz in his derivation of the lattice power-counting
theorem \cite{R,R(massless)} except that we do not need a certain technical condition 
on the lattice propagators that he required. This is because we are able to prove the 
(extended) power-counting theorem in the one-loop case in Appendix A without invoking this 
condition. It remains to be seen whether or not our techniques can be extended to the general 
multi-loop case so as to establish the general lattice power-counting theorem without using 
Reisz's additional condition.

We remark that the infrared finiteness condition (infrared divergence degree $<0$ 
\cite{R(massless)}) was only needed here for the application of the lattice power counting 
theorem to $h(p,M)=\lim_{a\to0}\ (I(p,a)-I_0(p,M,a))$ in \S3. The other applications of the
power-counting theorem (and its extension) are to lattice integrals associated solely with 
$I_0(p,M,a)$ and these are manifestly free of infrared divergencies for $M>0$.

The general structural results of this paper provide the basis for rigorously and explicitly
establishing several expected foundational properties of lattice QCD: universality of the one-loop 
beta-function, and universality of   
anomalous dimensions of renormalized operators at one loop (in the usual case where the 
renormalization factors are logarithmically divergent). The case of the one-loop lattice QCD 
beta-function requires some further elaboration since the individual lattice Feynman diagrams
relevant for its calculation can have stronger divergencies (linear and quadratic).
However, thanks to lattice BRST symmetry and lattice hypercube symmetries these integrals
can always be combined to get logarithmically divergent ones \cite{R(NPB),Kawai},
to which the results of the present paper can then be applied. From this the lattice QCD one-loop 
beta-function is seen to coincide with the continuum one, as we will show
explicitly in a future publication.
Furthermore, we hope to prove the consistency of general lattice formulations of QCD with
continuum QCD at one loop; i.e., show that when the same renormalization conditions are imposed
the renormalized Green's functions of interest are the same in the lattice and continuum 
formulations. This is of course expected, and and is known to be the case for the specific 
lattice QCD formulations studies to date, but it should be shown to hold for general 
lattice formulation.
One way to view this issue is as an attempt to answer the following question: ``What are the
mildest and most general conditions that a lattice QCD formulation must satisfy in order
to be consistent with continuum QCD?''\footnote{D.A. thanks Prof. Peter Weisz for suggesting 
this perspective.}

An important practical issue in lattice QCD is the explicit evaluation of lattice Feynman
integrals; this is needed, e.g., for the lattice-continuum matching of renormalized 
operators that is required in order to extract physical predictions from lattice QCD
(see, e.g., \cite{Capitani}). The results of this paper provide a secure theoretical basis for 
a standard procedure used to perform such evaluations in the case of one-loop logarithmically 
divergent integrals: To evaluate $I(a)$ one takes an already known lattice integral $I_0(a)$ 
whose logarithmically divergent part coincides with that of $I(a)$ so that $I(a)-I_0(a)$ 
is a finite lattice integral which can be evaluated, e.g., numerically \cite{Capitani}. 
But in order to choose an appropriate $I_0(a)$ we need to know what the 
divergent part of $I(a)$ actually is. More precisely, we need to know for sure that the 
divergent part of the integral is given exclusively by a log term $f\log(a)$ (no term
$\sim(\log(a))^{1/3}$ or the like), and we need to know the factor $f$. The structural
result of this paper tells us both of these things for general logarithmically divergent
one-loop lattice integrals.

There are several directions in which one could attempt to extend the results of this paper: to
logarithmically divergent multi-loop lattice integrals, general divergent one-loop integrals, 
and multi-loop integrals in general. Also, in connection with the Symanzik improvement program
it is useful to determine the structure of the terms in $I(p,a)$ which vanish for $a\to0$.
The expected structure of general divergent lattice integrals, which provides a guide for 
attempting to extend the results of this paper, is the following 
\cite{Symanzik,Luscher-Weisz(pert)}:
\be
I(p,a)=a^{-\omega}\sum_{n=0}^{\infty}\sum_{m=0}^l c_{mn}(p)a^n(\log\,a)^m
\label{6.1}
\ee
Here $\omega$ is the divergence degree of the integral and $l$ is the number of loops.
In the one-loop case this can be expressed as
\be
I(p,a)=a^{-\omega}(f(p,a)\log(a)+g(p,a))
\label{6.2}
\ee
where $f(p,a)=f_0(p)+f_1(p)a+f_2(p)a^2+\dots$ and $g(p,a)=g_0(p)+g_1(p)a+g_2(p)a^2+\dots$.

Finally, we mention that although naive and staggered lattice fermions do not satisfy the 
doubler-free condition ((iii) in \S2) it is nevertheless often possible to apply the results of 
this paper to lattice integrals of interest involving these. This can be done when the symmetries
of the integrand allow the integral to be rewritten as 
$\int_{-\pi/a}^{\pi/a}d^4k\,(\cdots)=N_t\int_{-\pi/2a}^{\pi/2a}d^4k\,(\cdots)$ where $N_t$
is the number of tastes ($=16$ for naive fermions and $4$ for staggered 
fermions).\footnote{For example, this is possible for the naive or staggered fermion loop 
contribution to the one-loop vacuum polarization \cite{Weisz}.}
The arguments and results of this paper are then easily carried over to the integral
$\int_{-\pi/2a}^{\pi/2a}d^4k\,(\cdots)$.

\begin{acknowledgements}

This research is supported by the KICOS international cooperative
research program (KICOS grant K20711000014-07A0100-01410), by the KOSEF grant 
R01-2003-000-10229-0), by the KRF grant KRF-2006-312-C00497, by the BK21 program of 
Seoul National 
University, and by the DOE SciDAC-2 program.

\end{acknowledgements}

\appendix

\section{Extended power-counting theorem for one loop lattice Feynman integrals}

For convenience we will treat $p$ and $a$ as dimensionless parameters here; i.e., we assume that
they are expressed in units of some arbitrary mass scale.

\noindent {\em Theorem.} For one-loop lattice integrals $I(p,a)$ satisfying the conditions
stated in \S2, if $\hd_I<0$ then the corresponding continuum integral $I^{(c)}(p)$ 
given by (\ref{2.10x}) is convergent and there exist $a_0(p)>0$ and $c(p)>0$ such that
\be
|I(p,a)-I^{(c)}(p)|\ \le\ c(p)\,a\log(1/a)\qquad\mbox{for $0<a\le a_0(p)$}\,.
\label{2.8}
\ee

{\em Remarks}. (i) The theorem implies the usual convergence statement 
$I(p,a)\stackrel{a\to0}{\longrightarrow}I^{(c)}(p)$ when $\hd_I<0$, which is Reisz's result
\cite{R,R(massless)} specialized to the one-loop case. The new content is the estimate on
how quickly the convergence happens. This plays an essential role in establishing the structural 
result of this paper since it is needed to prove the finiteness of $g(p)$ in \S4.
(ii) In practice it is often possible to obtain a sharper estimate than (\ref{2.8}). This will
be seen in the proof below -- see (\ref{2.p17}). (iii) The estimate (\ref{2.8}) is in accordance
with the general expectation for the lattice spacing dependence of $I(p,a)$ in 
Ref.\cite{Symanzik,Luscher-Weisz(pert)} 
which indicates that $I(p,a)-I^{(c)}(p)$ should vanish at least as fast
as $\sim a\log(a)$ for $a\to0$.

The convergence of the continuum integral $I^{(c)}(p)$ is a straightforward consequence of 
conditions (ii) in \S2 and the fact mentioned there that $d_P\le\hd_V$. We now make some technical
preparations for the proof of the rest of the theorem.

As noted in \S2 of \cite{R}, the difference between $V(k,p,a)$ and its continuum limit $P(k,p)$
admits an estimate of the form
\be
|V(k,p,a)-P(k,p)|\ \le\ a^{l_V}\,\sum_{j\in J} |Q_j(k,p)|\qquad\ l_V\ge1\ ,\ \ |k|\le\frac{\pi}{a}
\label{2.10}
\ee
where $J$ is a finite set and the $Q_j$'s are polynomials satisfying 
$d_{Q_j}\,\le\,\hd_V+l_V$. (This is readily seen by Taylor-expanding $F(ak,ap)$ in (\ref{2.2})
in powers of $a$.) An easy consequence of this is that for fixed $p$ there exist
constants $c_V(p)>0$ and $c_V'(p)>0$ such that
\be
|\,V(k,p,a)-P(k,p)|\ \le\ a^{l_V}(\,c_V(p)\,|k|^{\hd_V+l_V}+c_V'(p))
\qquad\mbox{for $\;|k|\le\frac{\pi}{a}$}
\label{2.11}
\ee
A similar argument shows the existence of $l_C\ge1$, $c_C(p)>0$ and $c_C'(p)>0$ such that
\be
|\,C(k,p,a)-E(k,p)|\ \le\ a^{l_C}(\,|k|^{n+l_C}+c_C'(p)) 
\qquad\mbox{for $\;|k|\le\frac{\pi}{a}$}
\label{2.12}
\ee
where we have used the fact that $\hd_C\!=\!n$ (cf. \S2).

The other technical properties that we will need are summarized in the following:

\noindent {\em Lemma.} 
For fixed $p$ there exist $b>0$, $\alpha>0$, $\gamma>0$, $\e>0$ with $\e\le1$,
and $R\ge0$ with $R\le\frac{\pi}{b}\e$, all depending on $p$ but independent of $a$, such that,
for $0<a\le b$, \hfill\break
\noindent (a) $\quad E(k,p)\;\ge\;2\alpha|k|^n\quad$ for $\ |k|\ge R$ \hfill\break
\noindent (b) $\quad C(k,p,a)\;\ge\;\alpha(\,|k|^n-R/2)\quad$ for 
$\ R\le|k|\le\frac{\pi}{a}\e$ \hfill\break
\noindent (c) $\quad C(k,p,a)\;\ge\;\frac{1}{\gamma a^n}\quad$ 
for $\ k\in\lb-\frac{\pi}{a},\,\frac{\pi}{a}\rb^4\,$, $\ |k|\ge\frac{\pi}{a}\e$ \hfill\break
%
%

\noindent {\em Proof of the lemma.} The existence of an $\alpha>0$ and $R\ge0$ such that part
(a) of the lemma is satisfied is an immediate consequence of condition (ii) on $E(k,p)$ 
stated in \S2. Next, re-expressing the estimate (\ref{2.12}) as 
\be
|\,C(k,p,a)-E(k,p)|\ \le\ c\,|ak|^l\,|k|^n+a^lc'\quad\mbox{for $\ |k|\le\frac{\pi}{a}$}
\label{2.16}
\ee
where we have set $c=c_C(p)$, $\;c'=c_C'(p)$ and $l=l_C\,$, and 
defining\footnote{The condition $\e\le1$ is to ensure that the region $|k|\le\frac{\pi}{a}\e$
is contained within $[-\frac{\pi}{a},\frac{\pi}{a}]^4$. The condition 
$a_1\le\frac{\pi}{R}\e$ is to ensure that $R\le\frac{\pi}{a_1}\e$.}
\be
\e:=\min\,\Big\lbrace\,\frac{1}{\pi}\Big(\frac{\alpha}{c}\Big)^{1/l}\,,\,1\,\Big\rbrace
\quad,\qquad a_1:=\min\,\Big\lbrace\,\Big(\frac{\alpha R}{2c'}\Big)^{1/l}\,,\,
\frac{\pi\e}{R}\,\Big\rbrace
\label{2.17}
\ee
we obtain
\be
|\,C(k,p,a)-E(k,p)|\ \le\ \alpha(\,|k|^n+R/2)\qquad
\mbox{for $\ |k|\le\frac{\pi}{a}\e\ $, $\  0<a\le a_1$.}
\label{2.18}
\ee
Combining this with the inequality
\be
C(k,p,a)&\ge&E(p,k)-|\,C(k,p,a)-E(k,p)|\,,
\label{2.19}
\ee
and using part (a) of the lemma, leads to part (b) for any choice of $b>0$ with $b\le a_1$. 
To prove part (c) we define $\gamma$ by
\be
\frac{2}{\gamma}=\min\,\Big\lbrace\,G(k,0)\;\Big|\ k\in\lb-\pi,\,\pi\rb^4\ ,\ k\ge\pi\e\,
\Big\rbrace\,.
\nonumber
\ee
The conditions (i) and (iii) on $G(k,0)$ in \S2 ensure that this minimum is strictly positive,
i.e. $\gamma>0$. Continuity of $G(k,q)$ in $k$ and $q$ then implies that there exists an
$R'>0$ (depending on $\e$, and hence on $p$) such that\footnote{Here $q$ is a variable 
collectively representing all the external momenta and masses, and $|q|$ is its norm in the
total space of these parameters.}
\be
G(k,q)\ \ge\ \frac{1}{\gamma}\qquad\mbox{for $\ k\in\lb-\pi,\,\pi\rb^4\,$, $\ k\ge\pi\e\,$,
$\ |q|\le R'$}
\label{2.20}
\ee
Consequently,
\be
C(k,p,a)=\frac{1}{a^n}\,G(ak,ap)\ \ge\ \frac{1}{\gamma a^n}
\qquad\mbox{for $\ k\in\lb\frac{-\pi}{a},\,\frac{\pi}{a}\rb^4\,$, $\ k\ge\pi\e\,$, 
$\ a\le\frac{R'}{|p|}$}
\label{2.21}
\ee
The lemma is hereby seen to hold with $b=\min\lbrace\,a_1\,,\,\frac{R'}{|p|}\rbrace$.

Parts (b) and (c) of the lemma are variants of bounds that arise in Reisz's approach
from an additional technical condition on the lattice propagators \cite{R}. 
Here we have derived the 
bounds using only the general conditions (i), (ii) and (iii) of \S2, without the need for Reisz's
additional condition. Our bounds suffice to prove the lattice power counting theorem in
the one loop case, as we will see below. It may be possible to modify Reisz's proof of
the general (arbitrary loop) lattice power counting theorem so that the present bounds 
also suffice there, without the need for the additional condition on the propagators,
but this is left for future work.

With the technical preparations in place we now proceed to the main goal:

\noindent {\em Proof of the theorem.} Choose $b$, $\alpha$, $\gamma$, $\e$, $R$ as in the lemma,
and restrict to lattice spacings $a\le b$. We rewrite the difference between the lattice and 
continuum integrals as follows:
\be
I(p,a)-I^{(c)}(p)&=&\int_{-\pi/a}^{\pi/a}d^4k\,\frac{V(k,p,a)}{C(k,p,a)}
-\int_{-\infty}^{\infty}d^4k\,\frac{P(k,p)}{E(k,p)} \nonumber \\
&=&\int_{|k|\le\frac{\pi}{a}\e}d^4k\,\bigg\lb\,\frac{V(k,p,a)}{C(k,p,a)}
-\frac{P(k,p)}{E(k,p)}\bigg\rb \nonumber \\
&&+\int_{k\in\lb-\frac{\pi}{a},\frac{\pi}{a}\rb^4\ ,\ |k|\ge\frac{\pi}{a}\e}
d^4k\,\frac{V(k,p,a)}{C(k,p,a)} \nonumber \\
&&-\int_{|k|\ge\frac{\pi}{a}\e}d^4k\,\frac{P(k,p)}{E(k,p)}
\label{2.p1}
\ee  
Splitting up the integration region $|k|\le\frac{\pi}{a}\e$ into the sub-regions
\be
{\cal R}_1&:&\ |k|\le R \nonumber \\
{\cal R}_2&:&\ R\le|k|\le\frac{\pi}{a}\e \nonumber
\ee
and rewriting the corresponding integrand as
\be
\frac{V(k,p,a)}{C(k,p,a)}-\frac{P(k,p)}{E(k,p)}
&=&\frac{V(k,p,a)-P(k,p)}{C(k,p,a)}
+P(k,p)\bigg\lb\,\frac{E(k,p)-C(k,p,a)}{E(k,p)C(k,p,a)}\bigg\rb \nonumber \\
&\equiv&{{\cal I}}_1(k,p,a)+{{\cal I}}_2(k,p,a)
\label{2.p1a}
\ee
we define 
\be I_{ij}:=\int_{{\cal R}_i}d^4k\,{\cal I}_j(k,p,a)\qquad\quad i,j\in\lbrace1,2\rbrace
\label{2.p1b}
\ee
Define $I_3(p,a)$ and $-I_4(p,a)$ to be the 2nd and 3rd integrals, respectively, 
in (\ref{2.p1}). Then $I(p,a)-I^{(c)}(p)$ equals the sum of these integrals, hence
\be
|I(p,a)-I^{(c)}(p)|&\le&|I_{11}(p,a)|+|I_{12}(p,a)|+|I_{21}(p,a)|+|I_{22}(p,a)| \nonumber \\
&&+|I_3(p,a)|+|I_4(p,a)|
\label{2.p3}
\ee
To prove the bound on $|I(p,a)-I^{(c)}(p)|$ stated in the theorem, it now suffices to establish
a similar bound on each term in the right-hand side of (\ref{2.p3}), which is what we do in
the following.

For $I_{11}(p,a)$ we use the bound (\ref{2.10}) to obtain
\be
|I_{11}(p,a)|&\le&\int_{|k|\le R}d^4k\,\frac{|V(k,p,a)-P(k,p)|}{C(k,p,a)} \nonumber \\
&\le&a^{l_V}\int_{|k|\le R}d^4k\,\frac{\sum_{j\in J}|Q(k,p)|}{C(k,p,a)}
\label{2.p4}
\ee
The infrared finiteness condition \cite{R(massless)} ensures that the integral here is finite
and remains so in the $a\to0$ limit (given by replacing $C(k,p,a)$ by $E(k,p)$ in the integrand). 
We conclude that $|I_{11}(p,a)|$ vanishes at least as fast as $\sim a^{l_V}$ for $a\to0$.

Applying a similar argument to $I_{12}(p,a)$ we find that
$|I_{12}(p,a)|$ vanishes at least as fast as $\sim a^{l_C}$ for $a\to0$.

For $I_{21}(p,a)$ we use the bounds in (\ref{2.11}) and part (b) of the lemma to obtain
\be
|I_{21}(p,a)|&\le&\int_{R\le|k|\le\frac{\pi}{a}\e}d^4k\,
\frac{|V(k,p,a)-P(k,p)|}{C(k,p,a)} \nonumber \\
&\le&\frac{a^{l_V}}{\alpha}\int_{R\le|k|\le\frac{\pi}{a}\e}d^4k\,
\bigg(\,\frac{c_V\,|k|^{\hd_V+l_V}+c_V'}{|k|^n-\frac{1}{2}R}\bigg)
\label{2.p6}
\ee
The $a\to0$ behavior of the last integral is
\be
\sim\,\int^{1/a}dr\,r^{3+\hd_V+l_V-n}
\label{2.p7}
\ee
Recalling (\ref{2.5}) we have $3+\hd_V+l_V-n=\hd_I+l_V-1$ where 
$\hd_I$ ($<0$) is the divergence degree of the lattice Feynman integral in the theorem.
It follows that for $a\to0$ the last integral in the right-hand side of (\ref{2.p6})
is convergent if $\,\hd_I+l_V\,<\,0$, diverges $\sim \log(1/a)$ if $\,\hd_I+l_V=0$, and 
diverges $\sim (1/a)^{\hd_I+l_V}$ if $\,\hd_I+l_v\,>\,0$.
Thus we obtain the following upper bounds on the rate
at which $I_{21}(p,a)$ vanishes for $a\to0\,$:\footnote{Here and in the following we use 
the notation $|I(a)|\,\le\;\sim a^q$ to denote that $|I(a)|$ vanishes at least as fast as
$\sim a^q$ for $a\to0$, etc.}
\be
|I_{21}(p,a)|\ \le\ \left\{
\begin{array}{ll}
\sim a^{l_V} & \ \mbox{for $\ \hd_I\,<\,-l_V$}\\
\sim a^{|\hd_I|}\log(1/a) & \ \mbox{for $\ \hd_I=-l_V$}\\
\sim a^{|\hd_I|} & \ \mbox{for $\ \hd_I\,>\,-l_V$}\\
\end{array}
\right.
\label{2.p8}
\ee

To derive a bound on $I_{22}(p,a)$ we first note that, since $d_P\le\hd_V$ (cf. \S2),
the polynomial function $P(k,p)$ admits a bound of the form
\be
|P(k,p)|\ \le\ c_P|k|^{\hd_V}+c_P'\qquad\forall\,k\in{\bf R}^4
\label{2.p9}
\ee
where the dependence of $c_P>0$ and $c_P'>0$ on $p$ has been suppressed in the notation.
Using this together with the bounds in (\ref{2.12}) and parts (a) and (b) of the lemma, 
(and the fact that $C(k,p,a)$ and $E(k,p,a)$ are positive by condition (i) of \S2) we obtain
\be
|I_{22}(p,a)|&\le&\int_{R\le|k|\le\frac{\pi}{a}\e}d^4k\,
|P(k,p)|\frac{|E(k,p)-C(k,p,a)|}{E(k,p)C(k,p,a)} \nonumber \\
&\le&\frac{a^{l_C}}{2\alpha^2}\int_{R\le|k|\le\frac{\pi}{a}\e}d^4k\,
\frac{(c_P|k|^{\hd_V}+c_P')(c_C|k|^{n+l_C}+c_C')}{|k|^n(|k|^n-\frac{1}{2}R)}
\label{2.p10}
\ee
The $a\to0$ behavior of the last integral is
\be
\sim\,\int^{1/a}dr\,r^{3+\hd_V+l_C-n}
\label{2.p11}
\ee
which is the same as (\ref{2.p7}) with $l_V$ replaced by $l_C$. By the same argument as in the 
preceding case we find the following upper bounds on the rate at which $I_{22}(p,a)$ vanishes
for $a\to0\,$:
\be
|I_{22}(p,a)|\ \le\ \left\{
\begin{array}{ll}
\sim a^{l_C} & \ \mbox{for $\ \hd_I\,<\,-l_C$}\\
\sim a^{|\hd_I|}\log(1/a) & \ \mbox{for $\ \hd_I=-l_C$}\\
\sim a^{|\hd_I|} & \ \mbox{for $\ \hd_I\,>\,-l_C$}\\
\end{array}
\right.
\label{2.p12}
\ee

To derive a bound on $I_3(p,a)$ we use the bounds in (\ref{2.p9}), (\ref{2.11}), and part (c)
of the lemma, and the fact that 
$k\in[-\frac{\pi}{a},\,\frac{\pi}{a}]^4\,\Rightarrow\,|k|\le\frac{2\pi}{a}\,$, to derive a bound
on the integrand:
\be
\frac{|V(k,p,a)|}{C(k,p,a)}&\le&\frac{|V(k,p,a)-P(k,p)|+P(k,p)}{C(k,p,a)} \nonumber \\
&\le&\frac{a^{l_V}\Big(c_V\Big(\frac{2\pi}{a}\Big)^{\hd_v+l_V}+c_V'\Big)
+c_P\Big(\frac{2\pi}{a}\Big)^{\hd_V}+c_P'}{1/\gamma a^n} \nonumber \\
&=&a^{n-\hd_V}\gamma\Big(c_V(2\pi)^{\hd_V+l_V}+a^{\hd_V+l_V}\,c_V'+c_P(2\pi)^{\hd_V}
+a^{\hd_V}c_P'\Big) \nonumber \\
&\equiv&a^{n-\hd_V}\chi(p,a)
\label{2.p13}
\ee
where $\chi(p,a)$ has a finite limit for $a\to0$. It follows that
\be
|I_3(p,a)|&\le&\int_{k\in\lb-\frac{\pi}{a},\frac{\pi}{a}\rb^4\ ,\ |k|\ge\frac{\pi}{a}\e}
d^4k\,\frac{|V(k,p,a)|}{C(k,p,a)} \nonumber \\
&\le&\Big(\frac{2\pi}{a}\Big)^4a^{n-\hd_V}\chi(p,a) \nonumber \\
&=&(2\pi)^4a^{|\hd_I|}\chi(p,a)
\label{2.p14}
\ee
showing that $|I_3(p,a)|$ vanishes at least as fast as $\sim a^{|\hd_I|}$ for $a\to0$.

Finally, for $I_4(p,a)$ we use the bounds in (\ref{2.p9}) and part (a) of the lemma to obtain
\be
|I_4(p,a)|&=&\int_{|k|\ge\frac{\pi}{a}\e}d^4k\,\frac{P(k,p)}{E(k,p)} \nonumber \\
&\le&\int_{|k|\ge\frac{\pi}{a}\e}d^4k\,\frac{c_P|k|^{\hd_V}+c_P'}{2\alpha|k|^n}
\label{2.p15}
\ee
The behavior of the last integral for $a\to0$ is
\be
\sim\,\int_{1/a}^{\infty}dr\,r^{3+\hd_V-n}
\label{2.p16}
\ee
Since $3+\hd_V-n=\hd_I-1$ it follows that the behavior is $\sim (1/a)^{\hd_I}=a^{|\hd_I|}\,$,
hence $|I_4(p,a)|$ vanishes at least as fast as $\sim a^{|\hd_I|}$ for $a\to0$.

From (\ref{2.p3}) and the $a\to0$ behaviors of the integrals derived above we obtain the
following upper bounds on the rate at which $I(p,a)$ converges to $I^{(c)}(p)$ for $a\to0\,$:
Set $l=\min\{l_V,l_C\}$, then
\be
|I(p,a)-I^{(c)}(p)|\ \le\ \left\{
\begin{array}{ll}
\sim a^l & \ \mbox{for $\ l\,<\,|\hd_I|$}\\
\sim a^{|\hd_I|}\log(1/a) & \ \mbox{for $\ l=|\hd_I|$}\\
\sim a^{|\hd_I|} & \ \mbox{for $\ l\,>\,|\hd_I|$}\\
\end{array}
\right.
\label{2.p17}
\ee
Since $l_V\ge1$, $\,l_C\ge1$, and $\hd_I\le-1$ we see that in all cases the convergence is 
at least as fast as $\sim a\log(1/a)$. This completes the proof of the theorem.

\section{Structure of $I_0^{(c)}(p,M)$ with dimensional regularization}

We consider the dimensionally regularized continuum integral $I_0^{(c)}(p,M)$:
\be
I_0^{(c)}(p,M,\e)=\int_{-\infty}^{\infty}l^{-\e}d^{4-\e}k\,\frac{P_0(k,p)}{E(k,0)+M^n}
\label{BB1}
\ee
where $l$ is an inverse mass parameter introduced for dimensional reasons. Our goal here is
to show that it has the form 
\be
I_0^{(c)}(p,M,\e)=f(p)\Big(\frac{-(lM)^{-\e}}{\e}\Big)+g^{(c)}(p)
\label{BB2}
\ee
up to terms which vanish for $\e\to0$, where $f(p)$ is the same factor that multiplies
the log-term in the corresponding lattice integral. This implies that the dimensionally
regularized continuum integral $I^{(c)}(p,\e)$ has the form claimed in the Introduction
with $f^{(c)}(p)=f(p)$.

In the calculations that follow
we repeatedly exploit the facts that $P_0(k,p)$ is homogeneous of order $m\!-\!r$ in $k$ 
and order $r$ in $p$; that $E(k,0)$ is homogeneous of order $n$ in $k$, and that 
$4+m-r-n=0$ (i.e., $\hd_I=0$).
Changing integration variable to $\hk=k/M$ in (\ref{BB1}) we find
\be
I_0^{(c)}(p,M,\e)=(lM)^{-\e}\int_{-\infty}^{\infty}d^{4-\e}\hk\,\frac{P_0(\hk,p)}{E(\hk,0)+1} 
\label{BB3}
\ee
Introducing a parameter $\lambda$ and changing variables to $\tk=\hk/\lambda$ we rewrite 
the integral as
\be
I_0^{(c)}(p,M,\e)&=&(lM)^{-\e}\int_{-\infty}^{\infty}d^{4-\e}\tk\,\lambda^{4-\e}\,
\frac{P_0(\lambda\tk,p)}{E(\lambda\tk,0)+1} \nonumber \\
&=&-\frac{(lM)^{-\e}}{\e}\int_{-\infty}^{\infty}d^{4-\e}\tk\,
\Big(\lambda\frac{d}{d\lambda}\lambda^{-\e}\Big)
\frac{\lambda^4P_0(\lambda\tk,p)}{E(\lambda\tk,0)+1} \nonumber \\
&=&-\frac{(lM)^{-\e}}{\e}\bigg\lb\,\lambda\frac{d}{d\lambda}\int_{-\infty}^{\infty}d^{4-\e}\tk\,
\lambda^{4-\e}\frac{P_0(\lambda\tk,p)}{E(\lambda\tk,0)+1} \nonumber \\
&&\qquad\qquad\quad-\ \int_{-\infty}^{\infty}d^{4-\e}\tk\,\lambda^{1-\e}\frac{d}{d\lambda}
\Big(\frac{P_0(\tk,p)}{E(\tk,0)+\lambda^{-n}}\Big)\,\bigg\rb
\label{BB4}
\ee
The first term on the right-hand side vanishes, 
since the intergal in it is seen to be independent of $\lambda$
after changing the integration variable back to $\hk=\lambda\tk$. The second term is calculated
to give
\be
I_0^{(c)}(p,M,\e)=\frac{(lM)^{-\e}}{\e}\,n\lambda^{-n-\e}\int_{-\infty}^{\infty}d^{4-\e}\tk\,
\frac{P_0(\tk,p)}{(E(\tk,0)+\lambda^{-n})^2}
\label{BB5}
\ee
Changing integration variable to $\hk=\lambda\tk$ leads to
\be
I_0^{(c)}(p,M,\e)=-\frac{(lM)^{-\e}}{\e}f^{(c)}(p,e)
\label{BB6}
\ee
where
\be
f^{(c)}(p,\e)=-n\int_{-\infty}^{\infty}d^{4-\e}\hk\,\frac{P_0(\hk,p)}{(E(\hk,0)+1)^2}
\label{BB7}
\ee
In the $\e\to0$ limit this reduces to the expression (\ref{3.7}) for $f(p)$. 
Hence $f(p,\e)=f(p)+O(\e)$, implying the claimed structure (\ref{BB2}).

\end{document}